\documentclass[twocolumn,runningheads,natbib]{svjour2}
\smartqed  
\usepackage{graphicx}
\journalname{Astrophysics and Space Science}
\begin{document}

\title{ Exotic bulk viscosity and its influence on neutron star r-modes
}

\author{Debarati Chatterjee         \and
        Debades Bandyopadhyay
}

\institute{D. Chatterjee \at
              Saha Institute of Nuclear Physics, 1/AF Bidhannagar, Kolkata-700064, India \\
           \and
           D. Bandyopadhyay \at
              Saha Institute of Nuclear Physics, 1/AF Bidhannagar, Kolkata-700064, India \\
}

\date{Received: date / Accepted: date}

\maketitle

\begin{abstract}
We investigate the effect of exotic matter in particular, hyperon matter on 
neutron star properties such as equation of state (EoS), mass-radius 
relationship and bulk viscosity. Here we construct equations of state within 
the framework of a relativistic field theoretical model. As 
hyperons are produced abundantly in dense matter, hyperon-hyperon interaction 
becomes important and is included in this model. Hyperon-hyperon interaction
gives rise to a softer EoS which results in a smaller maximum mass neutron star
compared with the case without the interaction. Next we compute the coefficient
of bulk viscosity 
and the corresponding damping time scale due to the non-leptonic weak process 
including $\Lambda$ hyperons. Further, we investigate the 
role of the bulk viscosity on gravitational radiation driven r-mode instability
in a neutron star of given mass and temperature and find that the instability 
is effectively suppressed.
\end{abstract}

\section{Introduction}
\label{intro}
The investigation of spin frequencies from burst oscillations of eleven 
nuclear-powered millisecond pulsars showed that the spin distribution had 
an upper limit at 730 Hz \citep{Cha2,Cha}. The fastest rotating neutron star 
discovered recently has a spin frequency 716 Hz \citep{Hes}.  In this respect, 
the study of r-modes in rotating neutron stars has generated 
great interest in understanding the absence of very fast rotating neutron stars
in nature. The r-modes are subject to 
Chandrasekhar-Friedman-Schutz gravitational radiation instability in rapidly
rotating neutron stars \citep{And98,And01,Fri98,Lin98,And99,Ster} which may
play an important role in regulating spins of young neutron 
stars as well as old, accreting neutron stars in low mass x-ray binaries 
(LMXBs). Another aspect of the r-mode instability is that this
could be a possible source for gravitational radiation \citep{And01,Gon,Nar} 
which may shed light on the interior of a neutron star.

It was realised that the r-mode instability could be effectively suppressed by 
bulk viscosity due to exotic matter in neutron star interior when the compact 
star cools down to a temperature $\sim {10}^9$ K. The coefficient 
of bulk viscosity due to non-leptonic weak processes involving exotic matter 
such as hyperon and quark matter was calculated by several
authors \citep{Jon1,Lin02,Pap3,Dal,Dra,Mad92,Mad00}. It was noted that 
relaxation times of the
non-leptonic processes were within a few orders magnitude of the period of 
perturbations and gave rise to large bulk viscosity coefficient. This led to 
the complete suppression of the r-mode instability.

In this paper, we investigate the effect of hyperon matter including 
hyperon-hyperon interaction on bulk viscosity coefficient and the r-mode 
stability. 
In Sec. 2 of this paper, we discuss the model used to calculate
equations of state, bulk viscosity coefficient and the corresponding 
time scale. Results of our calculation are
analysed in Sec. 3. Section 4 gives the outlook. 


\section{Model}
\label{sec:1}

\subsection{Equation of State}
We describe the 
$\beta$-equilibrated and charge neutral hyperon matter within the framework of
a relativistic field theoretical model where baryon-baryon interaction is 
mediated by the exchange of scalar and vector mesons and hyperon-hyperon 
interaction is taken into account by two strange mesons -- scalar $f_0$ 
(hereafter denoted as $\sigma^*$) and vector $\phi$ 
\citep{Sch93,Mis}. The Lagrangian density for hyperon matter including
non-interacting electrons and muons is written as
\begin{eqnarray}
{\cal L}_B &=& \sum_B \bar\Psi_{B}(i\gamma_\mu{\partial^\mu} - m_B
+ g_{\sigma B} \sigma + g_{\sigma^* B} \sigma^*\nonumber\\
&-& g_{\omega B} \gamma_\mu \omega^\mu
- g_{\phi B} \gamma_\mu \phi^\mu
- g_{\rho B}
\gamma_\mu{\mbox{\boldmath t}}_B \cdot
{\mbox{\boldmath $\rho$}}^\mu)\Psi_B\nonumber\\
&& + \frac{1}{2}\left( \partial_\mu \sigma\partial^\mu \sigma
- m_\sigma^2 \sigma^2\right) - U(\sigma) \nonumber\\
&& + \frac{1}{2}\left( \partial_\mu \sigma^*\partial^\mu \sigma^*
- m_{\sigma^*}^2 \sigma^{*2}\right) \nonumber\\
&& -\frac{1}{4} \omega_{\mu\nu}\omega^{\mu\nu}
+\frac{1}{2}m_\omega^2 \omega_\mu \omega^\mu \nonumber\\
&& - \frac{1}{4}{\mbox {\boldmath $\rho$}}_{\mu\nu} \cdot
{\mbox {\boldmath $\rho$}}^{\mu\nu}
+ \frac{1}{2}m_\rho^2 {\mbox {\boldmath $\rho$}}_\mu \cdot
{\mbox {\boldmath $\rho$}}^\mu\nonumber\\ 
&&-\frac{1}{4} \phi_{\mu\nu}\phi^{\mu\nu}
+\frac{1}{2}m_\phi^2 \phi_\mu \phi^\mu~.
\end{eqnarray}
The Dirac bispinor $\Psi_B$ represents isospin multiplets for baryons $B$=$N$, 
$\Lambda$, $\Sigma$ and $\Xi$. Here vacuum baryon mass is $m_B$,
and isospin operator $t_B$ while $\omega_{\mu\nu}$,  
$\phi{\mu\nu}$,  and $\rho_{\mu\nu}$ are field strength tensors. Neutrons and
protons do not couple with $\sigma^*$ and $\phi$ mesons. The scalar
self-interaction term \citep{Bog} is given by
\begin{equation}
U(\sigma) = \frac{1}{3} g_2 \sigma^3 + \frac{1}{4} g_3 \sigma^4 ~.
\end{equation}
We perform the calculation in the 
mean field approximation. The total energy density and pressure are 
respectively given by 
\begin{eqnarray}
{\varepsilon}
&=& \frac{1}{2}m_\sigma^2 \sigma^2 
+ \frac{1}{3} g_2 \sigma^3 + \frac{1}{4} g_3 \sigma^4
+ \frac{1}{2}m_{\sigma^*}^2 \sigma^{*2}\nonumber\\ 
&& + \frac{1}{2} m_\omega^2 \omega_0^2 + \frac{1}{2} m_\phi^2 \phi_0^2 
+ \frac{1}{2} m_\rho^2 \rho_{03}^2 \nonumber\\
&& + \sum_B \frac{2J_B+1}{2\pi^2} 
\int_0^{k_{F_B}} (k^2+m^{* 2}_B)^{1/2} k^2 \ dk \nonumber\\
&& + \sum_l \frac{1}{\pi^2} \int_0^{K_{F_l}} (k^2+m^2_l)^{1/2} k^2 \ dk
\end{eqnarray}
and 
\begin{eqnarray}
P &=& - \frac{1}{2}m_\sigma^2 \sigma^2 - \frac{1}{3} g_2 \sigma^3 
- \frac{1}{4} g_3 \sigma^4 \nonumber\\
&& - \frac{1}{2}m_{\sigma^*}^2 \sigma^{*2} 
+ \frac{1}{2} m_\omega^2 \omega_0^2 + \frac{1}{2} m_\phi^2 \phi_0^2 
+ \frac{1}{2} m_\rho^2 \rho_{03}^2 \nonumber\\
&& + \frac{1}{3}\sum_B \frac{2J_B+1}{2\pi^2} 
\int_0^{k_{F_B}} \frac{k^4 \ dk}{(k^2+m^{* 2}_B)^{1/2}}\nonumber\\
&& + \frac{1}{3} \sum_l \frac{1}{\pi^2} 
\int_0^{K_{F_l}} \frac{k^4 \ dk}{(k^2+m^2_l)^{1/2}}~, 
\end {eqnarray}
where $k_{F_B}$, $J_B$, $I_{3B}$ and $n_B$ are Fermi momentum, spin and isospin 
projection and number density of baryon $B$ respectively. The last terms in 
energy density and pressure are due to leptons. The effective baryon mass is 
defined as $m_B^*=m_B - g_{\sigma B}\sigma - g_{\sigma^* B}\sigma^*$.
The charge neutrality and chemical equilibrium conditions are given by
Q = $\sum_B q_B n_B -n_e -n_\mu =0$ and
$\mu_i = b_i \mu_n - q_i \mu_e$
where $\mu_n$, $\mu_e$ and $\mu_i$ are respectively
the chemical potentials of neutrons, electrons and baryon $i$ 
and $b_i$ and $q_i$ are baryon and electric charge of baryon $i$.
The chemical potential of baryons $B$ is given by \citep{Deba} 
$\mu_{B} = (k^2_{F_{B}} + m_B^{* 2} )^{1/2} + g_{\omega B} \omega_0
+ g_{\phi B} \phi_0 + I_{3B} g_{\rho B} \rho_{03}$. 

In this
calculation we adopt nucleon-meson coupling constants of GM set \citep{Gle91}.
The coupling constants are obtained by reproducing properties of saturated
nuclear matter 
such as binding energy $-16.3$ MeV, saturation density n$_0$ = 0.153 $fm^{-3}$, 
asymmetry
energy 32.5 MeV, effective nucleon mass 0.78 and incompressibility 240 MeV. 
The hadronic masses used in this calculation are $m_{\sigma}$ = 550 MeV, 
$m_{\omega}$ = 783 MeV, $m_{\rho}$ = 770 MeV and $m_N$ = 938 MeV.
On the other hand, hyperon-vector meson coupling constants are determined using 
SU(6) symmetry of the quark model \citep{Mis,Dov,Sch94} and the scalar $\sigma$
meson coupling to hyperons is calculated from hyperon potential depths  
in normal nuclear matter such as $U_{\Lambda}^N (n_0)$ = $-30$ MeV 
\citep{Dov,Chr}, $U_{\Xi}^N (n_0)$ = $-18$ MeV \citep{Fuk,Kha} and
a repulsive potential depth for $\Sigma$ hyperons $U_{\Sigma}^N (n_0)$ = +30 
MeV \citep{Fri1,Fri2,Bart} as obtained from hypernuclei data. 
The hyperon-$\sigma^*$ coupling constants are determined from double 
$\Lambda$ hypernuclei data \citep{Sch93,Mis}. 

We obtain the EoS solving equations of motion along with the expression for 
effective baryon mass and charge neutrality and beta equilibrium constraints
\citep{Mis,Deba}. 

\begin{figure}
\centering
\includegraphics[width=.95\columnwidth]{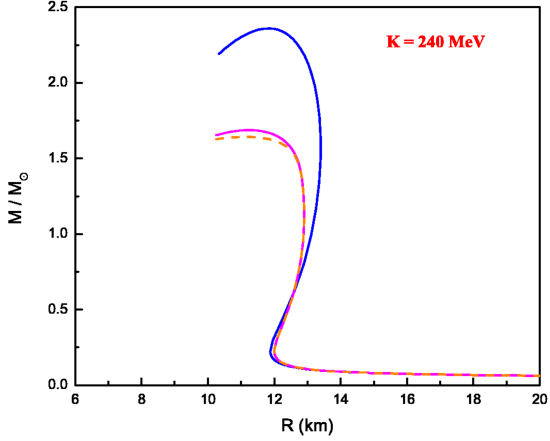}

\caption{ Mass-radius relationship of non-rotating neutron stars including
nucleons-only matter (blue solid line) and hyperon matter with (dashed line) 
and without (magenta solid line) hyperon-hyperon interaction.
}
\label{fig:1}       
\end{figure}

\subsection{Bulk viscosity coefficient, damping time scales and critical 
angular velocity}
Now we discuss the bulk viscosity coefficient in young neutron stars which 
cool down to temperatures $\sim$ $10^9-10^{10}$ K after their births. 
As the system goes out of chemical equilibrium due to pressure and density 
variations associated with the r-mode oscillations,
microscopic reaction processes in particular non-leptonic weak interaction 
processes including exotic particles might restore the equilibrium \citep{Nar,
Jon1,Lin02,Deba}. Here we calculate the real part of bulk viscosity 
coefficient ($\zeta$) in terms of relaxation times of microscopic processes 
\citep{Lan,Lin02} 
\begin{equation}
\zeta = \frac {P(\gamma_{\infty} - \gamma_0)\tau}{1 + {(\omega\tau)}^2}~,
\end{equation}
where the difference of infinite ($\gamma_{\infty}$) and zero ($\gamma_0$) 
frequency adiabatic indices is given by \citep{Lin02,Nar}  
\begin{equation}
\gamma_{\infty} - \gamma_0 = - \frac {n_b^2}{P} \frac {\partial P} 
{\partial n_n} \frac {d{\bar x}_n} {dn_b}~.
\end{equation}
In the co-rotating frame, the angular velocity 
($\omega$) of $(l,m)$ r-mode is 
related to angular velocity ($\Omega$) of a rotating 
neutron star as $\omega = {\frac {2m}{l(l+1)}} \Omega$ \citep{And01}. 
In this calculation, for $l=m=2$ r-mode, it is $\omega = {\frac {2}{3}} \Omega$.
The relaxation time ($\tau$) for the non-leptonic process
\begin{eqnarray}
n + p \rightleftharpoons p + \Lambda~,
\end{eqnarray}
is given by \cite{Lin02}

\begin{equation}
\frac {1}{\tau} = \frac {{(kT)}^2}{192{\pi}^3} {k_{F_{\Lambda}}}
{\langle {{|M_{\Lambda}|}^2} \rangle} \frac {\delta \mu}{{n_b}\delta{x_n}}~,
\end{equation}
where $k_{F_{\Lambda}}$ is the Fermi momentum for $\Lambda$ hyperons,
$\langle{|M_{\Lambda}|}^2 \rangle$ is the angle averaged squared matrix element 
and
\begin{equation}
\frac {\delta \mu}{{n_b}\delta{x_n}} = {\alpha_{nn}} - {\alpha_{\Lambda n}}
- {\alpha_{n \Lambda}} + {\alpha_{\Lambda \Lambda}}~,
\end{equation}
with 
$\alpha_{ij} = {\frac {\partial\mu_i}{\partial n_j}}_{{n_k},k \neq j}$. We 
obtain expressions for $\alpha_{ij}$ from the baryon chemical potential and
equations of motion for meson fields \citep{Deba}. For example,
\begin{eqnarray}
\alpha_{\Lambda n} &=& {\frac {\partial\mu_{\Lambda}}{\partial n_n}} 
 = {\frac {g_{\omega \Lambda} g_{\omega N}} {m_{\omega}^2}} \nonumber\\ 
&-& {\frac {m_{\Lambda}^* g_{\sigma \Lambda}}
{\sqrt{k_{F_{\Lambda}^2 + m_{\Lambda}^{*2}}}}} 
{\frac {\partial {\sigma}} {\partial n_n}} 
-{\frac {m_{\Lambda}^* g_{\sigma^* \Lambda}}
{\sqrt{k_{F_{\Lambda}^2 + m_{\Lambda}^{*2}}}}} 
{\frac {\partial {\sigma^*}} {\partial n_n}}~. 
\end{eqnarray}
Similarly, we compute other components of $\alpha_{ij}$. It is to be noted here
that nucleons do not couple with strange mesons, i.e. $g_{\sigma^* N}$ = 0.

The hyperon bulk viscosity damping timescale ($\tau_h$) contributes to the
imaginary part of the r-mode frequency and is calculated in the 
following way \citep{Nar,Lin98,Lin99} 
\begin{equation}
{\frac {1}{\tau_h}} =  - {\frac {1} {2E}} {\frac {dE}{dt}}~,
\end{equation}
where $E$ is the energy of the perturbation in the co-rotating frame
of the fluid 
\begin{equation}
E = \frac {1}{2}{\alpha^2}{\Omega^2}{R^{-2}} \int_0^R {\epsilon (r) r^6}dr~.
\end{equation}
The derivative of $E$ with respect to time is given by
\begin{equation}
\frac {dE}{dt} = -4 \pi \int_0^R \zeta (r) \langle |\vec{\nabla} \cdot 
{\delta \vec{v}}|^2 \rangle r^2 dr,~
\end{equation} 
where we need to know the energy density 
$\epsilon (r)$ and bulk viscosity $\zeta (r)$ profiles of a neutron star.
Similarly, we estimate the damping time scale ($\tau_U$) corresponding to 
the bulk viscosity due to the modified Urca process including only nucleons
using the bulk viscosity coefficient as given by \cite{Saw}. It was 
noted that gravitational radiation 
drives the r-modes unstable. So the gravitational radiation time scale 
($\tau_{GR}$) has a negative contribution to the imaginary part of the r-mode
frequency. Further, the overall r-mode time scale ($\tau_r$) is defined
as
\begin{equation}
{\frac {1}{\tau_r}} =  - {\frac {1}{\tau_{GR}}} + {\frac {1}{\tau_B}} + 
{\frac {1}{\tau_U}}.~ 
\end{equation}
The r-mode is stable below the critical angular velocity which is obtained 
from the solution of $\frac {1}{\tau_r}$ = 0. The critical angular velocity
depends both on the mass of the neutron star and its temperature.

\section{Results}
\begin{figure}
\centering
\includegraphics[width=.75\columnwidth,angle=90]{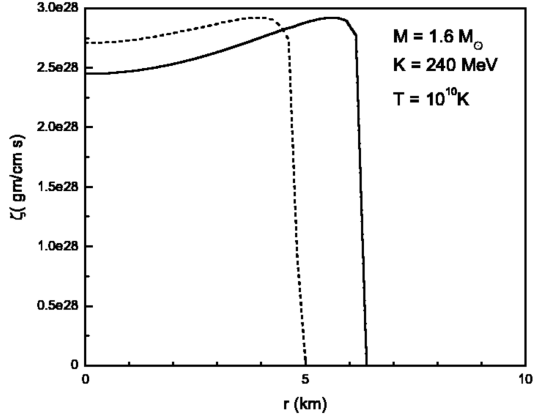}
\caption{ Bulk viscosity coefficient as a function of radial distance for a
non-rotating star (solid line) and as a function of equatorial distance for a 
rotating star (dashed line) of mass 1.6 
M$_{\odot}$ including hyperon matter with hyperon-hyperon interaction.
}
\label{fig:2}       
\end{figure}
We perform this calculation using the GM set \citep{Gle91} as it 
has been described in sec 2.
It is found that $\Lambda$ hyperons appear first at 2.6n$_0$ followed by $\Xi^-$
hyperons at 2.9n$_0$. However no $\Sigma$ hyperons appear in the system because
of strongly repulsive $\Sigma$-nuclear matter interaction \citep{Deba}. 
Further, we find that after the appearance of negatively charged $\Xi$ hyperons,
number densities of electrons and muons decrease in the system. This is 
attributed 
to the charge neutrality condition. Due to the appearance of additional degrees
of freedom in the form of hyperons, the EoS of hyperon matter becomes softer
compared with that of nucleons-only matter. On the other hand, the interplay 
between
the attractive $\sigma^*$ field and repulsive $\phi$ field which becomes 
dominant with increasing density, makes the EoS softer initially and stiffer 
at higher densities than the EoS without hyperon-hyperon interaction. 
This behaviour of the EoS is reflected in the calculation of 
maximum mass and the corresponding radius of the neutron star. In Fig. 1, we 
plot the mass-radius
relationship of non-rotating neutron stars calculated by solving 
Tolman-Oppenheimer-Volkoff equation for equations of state including
nucleons-only matter (blue solid line) and hyperon matter with (dashed line)
and without (magenta solid line) hyperon-hyperon interaction. Here the highest
maximum mass corresponds to the stiffest EoS of nucleons-only matter. The
maximum masses corresponding to the EoS with and without 
hyperon-hyperon interaction are 1.64 M$_{\odot}$ and 1.69 M$_{\odot}$
respectively. Using the rotating neutron star (RNS) model of Stergioulas 
\citep{Ster95}, we find that the maximum masses corresponding to the 
EoS with and without hyperon-hyperon interaction are respectively 1.95 
M$_{\odot}$ and 2.00 M$_{\odot}$.

Next we discuss the hyperon bulk viscosity due to the non-leptonic process 
given by Eq. (7). The bulk
viscosity coefficient is dependent on the EoS as it is evident from the 
expression in Eq. (5). The temperature dependence of the bulk viscosity enters
through the relaxation time as given by Eq. (8). The bulk viscosity coefficient
reaches a maximum and then drops with increasing baryon density as found by
\cite{Deba}. It has been also noted that the bulk viscosity increases as
temperature decreases. Therefore, the large value of $\zeta$ 
might be effective in suppressing r-mode instability 
as neutron stars cool down to a few times $10^{9}$ K.

\begin{figure}
\centering
\includegraphics[width=.75\columnwidth,angle=90]{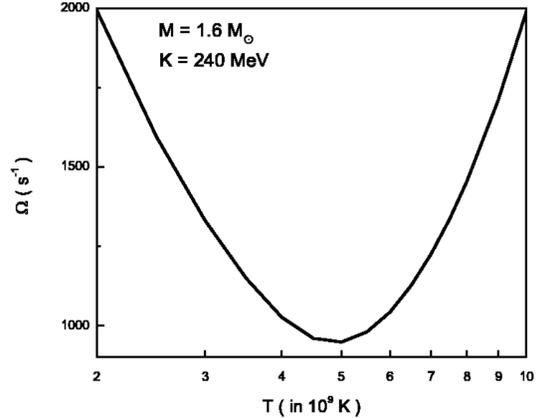}
\caption{ Critical angular velocity as a function temperature for a rotating
star of mass 1.6M$_{\odot}$ including hyperon matter with hyperon-hyperon 
interaction.
}
\label{fig:3}       
\end{figure}

The neutron star which was rotating rapidly at its birth,
slows down due to gravitational wave emission. 
A rotating neutron star has less central energy density than its 
non-rotating counterpart because of the 
centrifugal force. Consequently, hyperon thresholds are sensitive to 
rotation periods of compact stars. In the calculation of critical angular
velocity,  we consider a rotating neutron star of mass 1.6 M$_{\odot}$ 
and the corresponding
central baryon density is 3.9n$_0$ which is well above the threshold of 
$\Lambda$ hyperons. This star is rotating at an angular velocity 2952 s$^{-1}$.
It slowed down from its Keplerian angular velocity 5600 s$^{-1}$ when
the central baryon density was below the $\Lambda$ hyperon threshold. The
calculation of critical angular velocity requires the knowledge of energy 
density and bulk viscosity profiles in compact stars as it is evident from
Eqs. (11)-(14). 
In Fig. 2, the hyperon bulk viscosity profile (dashed line) of
the rotating neutron star of mass 1.6 M$_{\odot}$ is shown as a function of 
equatorial distance along with that of a  non-rotating (solid line) star of same
mass. We note that the peak of the profile shifts towards the center in the case
of the rotating neutron star. Earlier calculation of critical angular velocity
was performed using energy density and bulk viscosity profiles of non-rotating
neutron stars. Recently, \cite{Nar} studied the effect of rotation on critical
angular velocity. We calculate the critical angular 
velocity using energy density and bulk viscosity profiles of the rotating star.
In Fig. 3 the critical angular velocity is plotted versus
temperature. We find that the r-mode instability is suppressed by hyperon bulk 
viscosity. Here we do not consider the impact of superfluidity of 
baryons on bulk viscosity. However, several calculations 
\citep{Pap1,Pap2,Pap3,Nar} showed that superfluidity of baryons might
suppress the bulk viscosity and the damping of the r-mode instability. 

\section{Outlook}
Other forms of matter such as $K^-$ condensed matter and quark matter may 
appear and compete with the $\Lambda$ hyperon threshold in neutron star 
interior. In an earlier calculation we showed that the condensate of $K^{-}$ 
mesons appeared at 
$\sim$ twice normal nuclear matter density followed by the appearance of 
$\Lambda$ hyperons \citep{Bani}. It would be worth investigating how the
bulk viscosity coefficient due to the process (7) 
is modified in this situation. Also, we are studying whether the non-leptonic  
process involving $K^-$ condensate $n \rightleftharpoons p + K^-$ could give
rise to bulk viscosity coefficient as large as that of the non-leptonic process
(7) and damp the r-mode instability effectively \citep{CBS}. 

\begin{acknowledgements}
DB thanks the Alexander von Humboldt Foundation for support and 
acknowledges the warm hospitality at Frankfurt Institute for Advanced Studies
(FIAS) and Institute for Theoretical Physics, J.W. Goethe University,
Frankfurt am Main where a part of this work was completed.
\end{acknowledgements}


\begin{thebibliography}{3}
\bibitem[\protect\citeauthoryear{Andersson}{1998}]{And98}
Andersson N. ApJ, {\bf 502}, 708 (1998) 
\bibitem[\protect\citeauthoryear{Andersson, Kokkotas \& Schutz}{1999}]{And99}
Andersson N., Kokkotas K.D., Schutz B.F.  ApJ, {\bf 510}, 846 (1999) 
\bibitem[\protect\citeauthoryear{Andersson}{2003}]{And01}
Andersson N.  Class. Quant. Grav., {\bf 20}, R105 (2003)
\bibitem[\protect\citeauthoryear{Banik \& Bandyopadhyay}{2001}]{Bani}
Banik S.,  Bandyopadhyay D. Phys. Rev. C, {\bf 64}, 055805 (2001)
\bibitem[\protect\citeauthoryear{Bart et al.}{1999}]{Bart}
Bart S.et al. Phys. Rev. Lett., {\bf 83}, 5238 (1999)
\bibitem[\protect\citeauthoryear{Batty, Friedman \& Gal}{1997}]{Fri2}
Batty C.J., Friedman E., Gal A.  Phys. Rep., {\bf 287}, 385 (1997)
\bibitem[\protect\citeauthoryear{Boguta \& Bodmer}{1977}]{Bog}
Boguta J., Bodmer A.R.  Nucl. Phys., {\bf A292}, 413 (1977)
\bibitem[\protect\citeauthoryear{Chakrabarty et al.}{2003}]{Cha2}
Chakrabarty D. et al.  Nature, {\bf 424}, 42 (2003) 
\bibitem[\protect\citeauthoryear{Chakrabarty}{2005}]{Cha}
Chakrabarty D.  in the proceedings of Binary Radio
Pulsars, Rasio F. \& Stairs I., ASP Conference Series, Vol. 328, p.279 
\bibitem[\protect\citeauthoryear{Chatterjee \& Bandyopadhyay}{2006}]{Deba}
Chatterjee D., Bandyopadhyay D. Phys. Rev. D, {\bf 74}, 023003 (2006) 
\bibitem[\protect\citeauthoryear{Chatterjee, Bandyopadhyay \& Schaffner-Bielich}{2006}]{CBS}
Chatterjee D., Bandyopadhyay D., Schaffner-Bielich J. (in preparation) 
\bibitem[\protect\citeauthoryear{Chrien \& Dover}{1989}]{Chr}
Chrien R.E., Dover C.B.  Annu. Rev. Nucl. Part. Sci., {\bf 39}, 113 (1989)
\bibitem[\protect\citeauthoryear{Drago, Lavagno \& Pagliara}{2005}]{Dra}
Drago A., Lavagno A., Pagliara G.  Phys. Rev. D, {\bf 71}, 103004 (2005)
\bibitem[\protect\citeauthoryear{Dover \& Gal}{1984}]{Dov}
Dover C.B., Gal A.  Prog. Part. Nucl. Phys., {\bf 12}, 171 (1984)
\bibitem[\protect\citeauthoryear{Friedman, Gal \& Batty}{1994}]{Fri1}
Friedman E., Gal A., Batty C.J.  Nucl. Phys., {\bf A579}, 518 (1994) 
\bibitem[\protect\citeauthoryear{Friedman \& Morsink}{1998}]{Fri98}
Friedman J.L.,  S. M. Morsink  ApJ, {\bf 502}, 714 (1998) 
\bibitem[\protect\citeauthoryear{Fukuda et al.}{1998}]{Fuk}
Fukuda T. et al.  Phys. Rev. C, {\bf 58}, 1306 (1998)
\bibitem[\protect\citeauthoryear{Glendenning \& Moszkowski}{1991}]{Gle91}
Glendenning N.K., Moszkowski S.A. Phys. Rev. Lett., {\bf 67}, 2414 (1991)
\bibitem[\protect\citeauthoryear{Gondek-Rosinska, Gourgoulhon \& Haensel}{2003}]{Gon}
Gondek-Rosinska D., Gourgoulhon E., Haensel P.  A\&A, {\bf 412}, 777 (2003)
\bibitem[\protect\citeauthoryear{Haensel, Levenfish \& Yakovlev}{2000}]{Pap1}
Haensel P., Levenfish K.P., Yakovlev D.G. A\&A, {\bf 357}, 1157 (2000)
\bibitem[\protect\citeauthoryear{Haensel, Levenfish \& Yakovlev}{2001}]{Pap2}
Haensel P., Levenfish K.P., Yakovlev D.G. A\&A, {\bf 372}, 130 (2001)
\bibitem[\protect\citeauthoryear{Haensel, Levenfish \& Yakovlev}{2002}]{Pap3}
Haensel P., Levenfish K.P., Yakovlev D.G. A\&A, {\bf 381}, 1080 (2002)
\bibitem[\protect\citeauthoryear{Hessels et al.}{2006}]{Hes}
Hessels J.W.T. et al.  astro-ph/0601337 (2006)
\bibitem[\protect\citeauthoryear{Jones}{2001}]{Jon1}
Jones P.B.  Phys. Rev. Lett., {\bf 86}, 1384 (2001);
Phys. Rev. D, {\bf 64}, 084003 (2001)
\bibitem[\protect\citeauthoryear{Khaustov et al.}{2000}]{Kha}
Khaustov P. et al.  Phys. Rev. C, {\bf 61}, 054603 (2000)
\bibitem[\protect\citeauthoryear{Landau \& Lifshitz}{1999}]{Lan}
Landau L.D., Lifshitz E.M. Fluid Mechanics, Butterworth-Heinemann, Oxford 
(1999) 
\bibitem[\protect\citeauthoryear{Lindblom, Owen \& Morsink}{1998}]{Lin98}
Lindblom L., Owen B.J., Morsink S.M.  Phys. Rev. Lett.,  {\bf 80}, 4843 (1998) 
\bibitem[\protect\citeauthoryear{Lindblom, Mendell \& Owen}{1999}]{Lin99}
Lindblom L., Mendell G., Owen B.J. Phys. Rev. D, {\bf 60}, 064006 (1999)
\bibitem[\protect\citeauthoryear{Lindblom \& Owen}{2002}]{Lin02}
Lindblom L., Owen B.J.  Phys. Rev. D, {\bf 65}, 063006 (2002)
\bibitem[\protect\citeauthoryear{Madsen}{1992}]{Mad92}
Madsen J.  Phys. Rev. D, {\bf 46}, 3290 (1992)
\bibitem[\protect\citeauthoryear{Madsen}{2000}]{Mad00}
Madsen J. Phys. Rev. Lett., {\bf 85}, 10 (2000)
\bibitem[\protect\citeauthoryear{Nayyar \& Owen}{2006}]{Nar}
Nayyar M., Owen B.J. Phys. Rev. D, {\bf 73}, 084001 (2006)
\bibitem[\protect\citeauthoryear{Sawyer}{1989}]{Saw}
Sawyer R.F. Phys. Rev. D, {\bf 39}, 3804 (1989)
\bibitem[\protect\citeauthoryear{Schaffner et al.}{1993}]{Sch93}
Schaffner J. et al.  Phys. Rev. Lett., {\bf 71}, 1328 (1993)
\bibitem[\protect\citeauthoryear{Schaffner et al.}{1994}]{Sch94}
Schaffner J. et al. Ann. Phys. (N.Y.), {\bf 235}, 35 (1994)
\bibitem[\protect\citeauthoryear{Mishustin \& Schaffner}{1996}]{Mis}
Schaffner J., Mishustin I.N.  Phys. Rev. C, {\bf 53}, 1416 (1996)
\bibitem[\protect\citeauthoryear{Stergioulas \& Friedman}{1995}]{Ster95}
Stergioulas N., Friedman J.L. ApJ, {\bf 444}, 306 (1995)
\bibitem[\protect\citeauthoryear{Stergioulas}{2003}]{Ster}
Stergioulas N.  Liv. Rev. Rel., {\bf 6}, 3 (2003) 
\bibitem[\protect\citeauthoryear{van Dalen \& Dieperink}{2004}]{Dal}
van Dalen, E.N.E., Dieperink A.E.L.  Phys. Rev. C, {\bf 69}, 025802 (2004)
\end{thebibliography}
\end{document}